\documentclass[10pt,a4paper]{article}
\usepackage{jheppub}
\newcommand\ii{\'{\i}}
\usepackage{color}

\newcommand{\beq}{\begin{eqnarray}}
\newcommand{\eeq}{\end{eqnarray}}
\newcommand{\R}{\mathbb{R}}

\newcommand{\nn}{\nonumber}
\newcommand{\pa}{\partial}

\usepackage{epstopdf}

\title{Thermodynamics of conformal fields in topologically non-trivial space-time backgrounds}
\author[a]{M. Asorey,}
\author[b]{C. G. Beneventano,%
}
\author[b]{D. D'Ascanio%
}
\author[b]{and E. M. Santangelo}
\affiliation[a]{Departamento de F\ii sica Te\'orica, Universidad de Zaragoza\\
E-50009  Zaragoza, Spain }
\affiliation[b]{Departamento de F\ii sica, Universidad Nacional de La Plata\\
Instituto de F\ii sica de La Plata, CONICET-Universidad Nacional de La Plata\\
C.C.67, 1900 La Plata, Argentina}
\emailAdd{asorey@unizar.es}
\emailAdd{gabriela@fisica.unlp.edu.ar}
\emailAdd{dascanio@fisica.unlp.edu.ar}
\emailAdd{mariel@fisica.unlp.edu.ar}
\abstract{{We analyze the finite temperature behaviour of massless conformally coupled
scalar fields in homogeneous lens spaces $S^3/{\mathbb Z}_p$.
High and low temperature expansions
are explicitly computed and the behavior of thermodynamic quantities under thermal duality is scrutinized.
The analysis of the entropy in the high-temperature limit of the different lens spaces  points out the
appearance of a temperature-independent contribution of topological origin to the entropy. The remaining terms are
exponentially suppressed by the temperature. The topological contribution to the
entropy appears as a subleading correction to the Stefan-Boltzmann term of the free energy, and can be obtained from
the determinant of the lens space conformal Laplacian operator. In the low-temperature limit the leading term in the free energy is the Casimir
energy and there is no trace of any power correction in any lens space. In fact, the remaining corrections are always exponentially suppressed by the
inverse of the temperature. The duality between the results of both expansions is further analyzed in the paper.} }

\notoc
\begin{document}
\preprint{}

\maketitle
\vfill

\section{Introduction}

The appearance of black holes, the Unruh effect  and the discovery of Hawking radiation
pointed out, time ago, the thermodynamical behavior induced by  gravitational backgrounds. It is known that the existence of causal horizons generates the basic ingredients for a thermodynamic background: entropy and temperature. Space-time horizons in cosmological backgrounds also provide the basis
for the presence of a similar thermodynamics at cosmological scales \cite{gh}. A dramatic twist on this connection was
produced by the formulation of the holographic principle \cite{thooft, susskind} and the conjecture about the thermodynamical origin of the gravitational interaction \cite{jacob, erikv}. The thermodynamics induced by  gravitational backgrounds is very special, with expected bounds
on the entropy induced by quantum fields \cite{bekenstein, bekenstein2,bekenstein3, bousso}. In the case of closed manifolds, a particularly interesting point is the conjectured validity of the Cardy-Verlinde  formula \cite{cardy1,evcardy}. Such formula is a natural generalization
of  Cardy's formula, which holds for any 1+1 conformal field theory. The hint for the extension of Cardy's formula to higher
dimensions was the behavior of the holographic AdS/CFT correspondence in the strong coupling regime. Checks of the validity of the Cardy-Verlinde formula in the
high-temperature limit of free field theories were the subject of several papers \cite{kutasov,odintsov,sinopsys}. However, the results in some of these papers were not concluding, due to the presence of ambiguities associated with zero modes \cite{dowker}.

In 1+1 dimensions free bosonic field theories have zero modes and one way of removing their ambiguities is by compactification of  the field into a sigma model, as in string theory \cite{aj2}. To avoid ambiguities, in this paper we shall explore the dependence of the entropy on the topological structure of spherical 3-dimensional manifolds. In these cases, no zero modes appear and, thus, no ambiguities remain when evaluating thermodynamic quantities. In particular, we shall consider space-times with $\R\times L(p,1)$ topology, where $L(p,1)=S^3/{\mathbb Z}_{p}$ is a
compact three-dimensional homogeneous lens space.

In general, thermodynamic quantities are highly dependent on the geometric and topological properties of the spatial manifold.
We shall focus on the behaviour of the entropy. There are two types of interesting subleading corrections to the entropy which deserve study. In the case of compact manifolds with boundaries, bulk entropy is an extensive quantity, since it is proportional to the volume of the space. On the contrary, boundary entropy has not such an extensive character and might mimic the black-hole entropy behavior. In 1+1 dimensions, this entropy is dimensionless and can have a logarithmic dependence on the space volume or be a space independent quantity \cite{aj2,cardy,affleck,affleck2}. In a manifold without boundary, a similar dimensionless subleading contribution can arise and, because of its dependence on the topological structure of the space, it can be called a {\it topological  subleading entropy}. This contribution is reminiscent of the topological entanglement
entropy which appears in field theories when tracing out degrees of freedom on a space domain \cite{kitaev}.

The current cosmological model is consistent with a spatially flat Universe, although most of the relevant data are compatible with a very tiny curvature \cite{var,kom}. In this case, a possible space-time topology is $\R\times S^3$. However, the physical observations do not allow to establish in a conclusive way whether the space is simply or multiply connected. Closed spaces leave their fingerprints in small contributions to low multipoles of the Cosmic Microwave Background (CMB) and current observations do show a strong suppression of low multipoles (quadrupole, octupole, etc.). They also show a strange alignment of the quadrupole and octupole multipoles associated to the appearance of Southern hemisphere cool fingers. All these data suggest a possible role of the finite size and topology of the space manifold in the low-mode behavior of the CMB. Non-trivial topologies of space-time have been suggested as a possible
explanation of the presence of anomalies in the spectrum of CMB observed by WMAP \cite{cornish,luminet} (see also \cite{lachi,sokolov,zeldovich,fang,fagundes,acm}).
This is another motivation to explore possible space-time topologies of the form $\R\times L(p,1)$. In fact, space-times with non-trivial topology $L(p,1)$ were considered already by de Sitter
\cite{desitt}.

To analyze the thermodynamics of fields on such  manifolds we shall consider for simplicity a static Einstein metric, which can be obtained from the FLRW metric via a conformal mapping. The thermodynamics in static Einstein space-times was first studied by Dowker and Critchley \cite{DC} and, in de Sitter space-times, by Gibbons and Hawking \cite{gh}.

In this paper, we perform a complete evaluation of the relevant functional determinants, through the zeta function regularization technique \cite{zeta}. Using the duality transformation properties of the infinite sums appearing in the evaluation of the logarithm of the determinant, we obtain two equivalent expressions for the effective action on $S^3$ and on any homogeneous lens space, which provides the basis for the low- and high-temperature expansions of the free energy.

We focus on the analysis of subleading corrections to the thermodynamic quantities associated to conformal scalar quantum fields due to non-trivial background topologies.  Einstein space-times over lens spaces $L(p,1)$ have been already considered by Dowker and coworkers at zero temperature \cite{Dowker78,Dowker89,Dowker041,Dowker042}, and an approach to the finite temperature problem with a dynamical spatial metric in three Euclidean space-time dimensions, through squashed lens spaces, can be found in \cite{marcelo}.

In section \ref{high}  we give a first expression for the finite temperature effective action, which leads to a high-temperature expansion and show that, apart from the usual  Stefan-Boltzmann term, an extra, temperature independent term appears, which is the origin of a non-vanishing topological  subleading entropy. We also derive explicit expressions for exponential corrections to both contributions.

In section \ref{low} we give a second, equivalent, expression for the same object, which gives a low-temperature expansion and allows for the determination of the Casimir energies of all the spherical spaces considered. The remaining contributions are exponentially suppressed. In all cases, the internal energy does coincide with the Casimir energy and, thus, the entropy vanishes, in agreement with the first law of Thermodynamics.

Section \ref{thermo} contains the evaluation of the free and internal energies, as well as the entropy, in both temperature limits, together with an analysis of the behavior of the internal energies under the thermal duality (T-duality) transformation.

Finally, we rederive in appendix \ref{ap1} the effective action on the 3-sphere $S^3$ using the Abel-Plana formula in order to clarify some
confusing issues found in the literature.

\section{Conformal scalar field in $S^1\times L(p,1)$}

Let us consider for simplicity the case of a massless scalar field with conformal coupling. The action is given by
\beq
\begin{array}{lll}
S_{\mathrm c}  =  \displaystyle \frac12\,\int d^4x
\sqrt{g}\,&\Big\{&
               \displaystyle g^{\mu\nu}\pa_\mu \phi\,\pa_\nu \phi
                + \frac{n-1}{4n} R\phi^2
             \Big\}\,,
 \label{scalar action}
\label{C2}
\end{array}
\eeq
where $R=\frac{n(n-1)}{a^2}$ with $n=3$ is the curvature of $L(p,1)$.

The partition function can be computed by functional integral methods.
A finite temperature can be introduced by compactifying the Euclidean time and imposing
 periodic boundary conditions, according to the bosonic statistics  of scalar fields. This amounts
 to considering Euclidean functional integrals over $S^1\times L(p,1)$.
 The  functional integral associated to
the Euclidean action \eqref{C2} is a gaussian integral whose covariance operator is
\beq
-\square{}+ \frac{n-1}{4n} R= -\frac1{\sqrt{g}}\, \partial_\mu g^{\mu\nu}{\sqrt{g}}\,  \partial_\nu + \frac{(n-1)^2}{4 a^2 }.
\label{covariance}
\eeq

The $S^1$ modes are given by  Matsubara frequencies $\omega _l =\frac{2\pi l}{\beta}$ and the spatial modes are given by the eigenvalues of the Laplacian operator $-\Delta$ on $L(p,1)$ shifted by $\frac{1}{a^2}$.
This extra term comes from the conformal coupling to the curvature (last term in \eqref{scalar action}) and is essential to
 avoid the existence of zero modes and the presence of ambiguities in the derivation of thermodynamic quantities. Such ambiguities do affect similar calculations in 1+1 dimensions, where the conformal term vanishes.
The eigenvalues of the operator $-\Delta+\frac{1}{a^2}$
are given by
\beq
\lambda_k^2=\left(\frac{k}{a}\right)^2, \quad k=1,2,\ldots,\infty
\eeq
and their  degeneracy depends on the parity of $p$. For $p=2q+1$ odd the degeneracies  are
\beq
d_k= \begin{cases} k\left[\frac{k}{2q+1}\right],&\mbox{for}\quad k-(2q+1)\left[\frac{k}{2q+1}\right]\quad \mbox{even}\\
k\left(\left[\frac{k}{2q+1}\right]+1\right),&\mbox{for}\quad k-(2q+1)\left[\frac{k}{2q+1}\right]\quad \mbox{odd,}\end{cases}\,
\label{odd}
\eeq
 whereas, for $q$ even, they are
\beq
     d_k =\begin{cases} 0,&  \quad \mbox{for} \quad k \quad \mbox{even} \\
      k\left(2\left[\frac{k}{2q}\right]+1\right),& \quad \mbox{for} \quad  k \quad \mbox{odd.}\end{cases}\,
      \label{even}
\eeq
In both cases  $[x]$ denotes the integer part of $x$.

The finite temperature one--loop effective action can be evaluated, by using zeta-function regularization \cite{zeta}, as
\beq S_{\rm eff}=-\log { Z}_\beta\equiv-\frac12 \left.\frac{d}{ds}\zeta_M(s)\right\vert_{s=0}\,,\label{seff}\eeq
where
$$\zeta_M(s)={\mu}^{2s} \sum_{l=-\infty}^\infty\sum_{k=1}^{\infty} d_k\left(\left(\textstyle \frac{2\pi l }{\beta}\right)^2+\lambda_k^2 \right)^{-s}\,, \label{zeta}$$
$M=L(p,1)$ being the spatial manifold. As usual, the parameter $\mu$ is introduced to render the zeta function dimensionless and to take into account renormalization group dependences. As we will see, in our case the results will be  $\mu$--independent. This is so because we are dealing with a conformal classical theory and the conformal anomaly vanishes due to the special geometry of Einstein spaces and the special topology of $S^1\times L(p,1)$. This anomaly is given by
\beq
T=T^\mu_\mu\,=\,\frac{1}{1920 \pi^2} C^2 -\,\frac{1}{5760\pi^2}
E+a^\prime{\square}R,
\label{trace}
\eeq
where $R$ is the scalar curvature,
$$
\begin{array}{ll}
\displaystyle{
C_{\alpha\beta\mu\nu}}=&\displaystyle{R_{\alpha\beta\mu\nu}
+\frac{1}{2}\,(g_{\beta\mu}R_{\alpha\nu}-g_{\alpha\mu}R_{\beta\nu}
+g_{\alpha\nu}R_{\beta\mu}-g_{\beta\nu}R_{\alpha\mu})}
\phantom{\Big ]}\cr
&\displaystyle{ +\frac{1}{6}\,R\,(g_{\alpha\mu}g_{\beta\nu}-
g_{\alpha\nu}g_{\beta\mu})\,.}
\end{array}
$$
 the  Weyl conformal curvature tensor, and
$$
E = R_{\mu\nu\alpha\beta}R^{\mu\nu\alpha\beta}
- 4 \,R_{\alpha\beta}R^{\alpha\beta} + R^2\,.
$$
the Gauss-Bonet term. The coefficient $a'$ is ambiguous \cite{shapiro}, but since the curvature $R$ is constant the last term  of the anomaly (\ref{trace})  vanishes.  The Weyl tensor $C$ also vanishes because the FLRW metric is conformally flat. Finally, the  Gauss-Bonet term also vanish identically due to the special geometry of $S^1\times L(p,1)$. In fact, it is immediate to verify that the Euler-Poincar\'e characteristic $\chi=b_0-b_1+b_2-b_3+b_4$
of    $S^1\times L(p,1)$ is zero ($\chi=1-1+0-1+1=0$) for any  space-time metric;  and this
characteristic $\chi$ is nothing but the integral of $E$ on  $S^1\times L(p,1)$. However,
 in general, this is not the case and there is a  non-vanishing trace (conformal) anomaly in the four dimensional theory \cite{dowkerkenn,svaiter}.
 In such cases, a logarithmic dependence on $\mu$ remains.

\section{A first expression for $S_{\mathrm{eff}}$ and the high-temperature limit}
\label{high}

In this section, we will perform the analytic extension of the zeta function in a way which is adequate to isolate the infinite temperature ($\beta \rightarrow 0$) terms from their corrections. We will show that these last terms are exponentially suppressed  in all cases. The determination, through a different extension, of alternative expressions useful in the zero-temperature limit, will be obtained in the next section.

\paragraph{3-sphere $M=S^3$.}

In this case, the degeneracies are given by $d_k =k^2$. Thus we have

\begin{equation}
 \zeta_{S^3}(s)= ({\mu\,a})^{2s} \sum_{l=-\infty}^\infty\sum_{k=1}^{\infty} k^2 \left(\left(\frac{2\pi a l}{\beta}\right)^2+k^2 \right)^{-s}.
 \label{zetas3}
 \end{equation}

 Now, the previous expression can be written as $\zeta_{S^3}(s)=\zeta_{S^3}^{l=0}(s)+\zeta_{S^3}^{l\neq 0}(s)$, with
\beq
\zeta_{S^3}^{l=0}(s)= ({\mu\,a})^{2s}\sum_{k=1}^{\infty}k^{-(2s-2)}= ({\mu\,a})^{2s}{\zeta}_R (2s-2)\,,
\label{z03}
\eeq
where ${\zeta}_R (z)$ is Riemann's zeta function. We remark that the contribution of this term to the effective
action
\beq
-\frac12\left.\frac{d}{ds}\zeta_{S^3}^{l=0}(s)\right\vert_{s=0}=\frac12\log \det (-\Delta+\frac{1}{a^2}),
\eeq
does coincide with
the effective action of the same bosonic field on the Euclidean space $S^3$. This is a general feature which
 holds for all space times of the form $S^1\times M$. In particular it will also be true for lens spaces, where $M=L(p,1)$.

The higher-mode contribution can be splitted into two terms,
\beq
\zeta_{S^3}^{l\neq 0}(s)&=&({\mu\,a})^{2s} \left\{\sum_{l=1}^\infty\sum_{k=-\infty}^{\infty}\left[k^2+\left(\frac{2l\pi a}{\beta}\right)^2\right]^{-(s-1)}\!\!\!\!\!-\,\,\,\!\!\sum_{l=1}^\infty\left(\frac{2l\pi a}{\beta}\right)^2\sum_{k=-\infty}^{\infty}\left[k^2+\left(\frac{2l\pi a}{\beta}\right)^2\right]^{-s}\right\}\nn\\&=&\zeta_{1}(s)+\zeta_{2}(s)\,.
\eeq
We can write $\zeta_{1}(s)$ in terms of the Mellin transform
\beq
\zeta_{1}(s)=\frac{({\mu\,a})^{2s}}{\Gamma(s-1)}\sum_{l=1}^\infty\sum_{k=-\infty}^{\infty}\int_0^{\infty}dt\,t^{s-1-1}e^{-t\left[k^2+\left(\frac{2l\pi a}{\beta}\right)^2\right]}\,.
\eeq
The infinite sum over $k$ can be treated via the Poisson inversion formula for the Jacobi theta functions, or its equivalent

\beq\sum_{k=-\infty}^{\infty}e^{-t(k+c)^2}={\left(\frac{\pi}{t}\right)}^{\frac12}\sum_{k=-\infty}^{\infty}e^{-\left(\frac{{\pi}^2 k^2}{t}\right)}e^{2\pi i kc}\,,\label{kk}\eeq
which gives as a result
\beq\zeta_{1}(s)=\frac{({\mu\,a})^{2s}{\pi}^{\frac12}}{\Gamma(s-1)}\sum_{l=1}^\infty\sum_{k=-\infty}^{\infty}\int_0^{\infty}t^{s-\frac32-1}e^{-\frac{\pi^2\,k^2}{t}}e^{-t\left(\frac{2l\pi a}{\beta}\right)^2}\,.\eeq

After performing the integrals, and using $\Gamma(s-1)=\frac{\Gamma(s+1)}{(s-1)s}$, one finally has
\beq\zeta_{1}(s)&=&\frac{({\mu\,a})^{2s}(s-1)s{\pi}^{\frac12}}{\Gamma(s+1)}\left\{\Gamma(s-\frac32)\left(\frac{2\pi a}{\beta}\right)^{3-2s}{\zeta}_R (2s-3)\right.\nn\\
&+& \left. 2\sum_{l=1}^\infty\sum_{k=1}^{\infty}\left(\frac{2la}{k\beta}\right)^{\frac32-s}K_{s-\frac32}\left(\frac{4lka{\pi}^2}{\beta}\right)\right\}\,,
\label{z13}
\eeq
where $K_{\nu}(x)$ is the modified Bessel function of order $\nu$ \cite{rusa}.

The analytic extension of $\zeta_{2}(s)$ is performed in the same way and gives as a result
\beq\zeta_{2}(s)&=&-\frac{({\mu\,a})^{2s}s{\pi}^{\frac12}}{\Gamma(s+1)}\left\{\Gamma(s-\frac12)\left(\frac{2\pi a}{\beta}\right)^{3-2s}{\zeta}_R (2s-3)\right.\nn\\
&+& \left. 2\sum_{l=1}^\infty\sum_{k=1}^{\infty}\left(\frac{2l\pi a}{\beta}\right)^{2}\left(\frac{2la}{k\beta}\right)^{\frac12-s}K_{s-\frac12}\left(\frac{4lka{\pi}^2}{\beta}\right)\right\}\,.
\label{z23}
\eeq

Looking back at equations \eqref{z03}, \eqref{z13} and \eqref{z23} one can verify that, as commented before, $\zeta_{S^3}(0)=0$. For this reason it is easy to perform the $s$-derivative and, thus, to evaluate the effective action for the 3-sphere according to \eqref{seff}. After using the explicit expressions for the modified Bessel functions, the final result is
\beq
S_{\mathrm{eff},\,S^3}&=&\frac{1}{4{\pi}^2}{\zeta}_R (3)-\frac{{\pi}^4}{45} \left(\frac{a}{\beta}\right)^3\nn\\
 &+& \frac{1}{4\pi^2}\sum_{k,l=1}^{\infty}\frac{1}{k^3} \left[ 2 + 2\frac{4\pi^2kal}{\beta} + \left( \frac{4\pi^2kal}{\beta} \right)^2 \right] e^{-\frac{4\pi^2kal}{\beta}}.
\label{eff3}
\eeq

This expression is valid for all temperatures. In particular, in the high-temperature limit only the first two terms remain. The first term (which comes from $l=0$) is the one which gives rise to the topological  subleading entropy, while the second one is the volume term, with the usual Stefan-Boltzmann dependence on the temperature.

As already said, the topological contribution to the effective action \eqref{eff3} can be directly computed from the determinant of the corresponding spatial operator
\beq
\frac12\log \det (-\Delta_{S^3}+\frac1{a^2})= \frac{1}{4\pi^2}\zeta_R(3).
\eeq

Here, it is interesting to note that the infinite temperature limit for this same case was studied time ago in \cite{kutasov,emilio}. However, the constant term does not appear in any of these references. The origin of this absence is clear in the case of the first reference, where the authors evaluate the internal energy in this limit and disregard the arbitrary constant when integrating back to obtain the effective action. Why it is missing in the second case will be analyzed in appendix \ref{ap1}, where we shall also explore the connection between the topological subleading entropy and the extensive entropy density of the same theory in a two dimensional space.

\bigskip

As we will see in what follows, the knowledge of the effective action for the 3-sphere (the covering space) will greatly simplify the calculation in all the lens spaces $L(p,1)$. Because of the different behavior of the degeneracy
of the spectrum for even or odd values of $p$ we shall study both cases separately.

\paragraph{Odd lens spaces $M=L(2q+1,1)$.}
\label{ss22}

\par
As already mentioned, the knowledge of the complete effective action for $S^3$ proves extremely useful in simplifying the task of determining the effective action for these lens spaces.
The degeneracies of the eigenvalues $\lambda_k^2$ are given by \eqref{odd}. Now,
 writing $k=n(2q+1)+m$ with $n=0,1,...,\infty$ and $m=0,1,...,2q$, we obtain
\beq
d_k= \begin{cases}k\left(\frac{k-m}{2q+1}\right),&  \quad \mbox{for} \quad m \quad \mbox{even} \\
       k\left(\frac{k-m}{2q+1}+1\right),& \quad \mbox{for} \quad  m \quad \mbox{odd.}\end{cases}
\label{deg2q+1}\eeq
The first term in both lines, when summed over all possible values of $m$, will reproduce the degeneracy of $S^3$, divided by the order of the cyclic group. As a consequence,
\beq\zeta_{S^3/{\mathbb Z}_{2q+1}}(s)=\frac{\zeta_{S^3}(s)}{2q+1}+\delta\zeta_{S^3/{\mathbb Z}_{2q+1}}(s)\,.\label{2q+1}\eeq

Using equation \eqref{deg2q+1} we have
\beq
\delta\zeta_{S^3/{\mathbb Z}_{2q+1}}(s)&=&
-\frac{({\mu\,a})^{2s}}{2q+1}\sum_{\substack{l=-\infty\\ n=0}}^{\infty}\left\{\sum_{m=1}^q 2m\left(n(2q+1)+2m\right)\left[\left(n(2q+1)+2m\right)^2 +\left(\frac{2la\pi}{\beta}\right)^2\right]^{-s}\right. \nn \\  && \!\!\!\!\!\!\!\!\!\!\!\!\!\!\!\!\!\!\!\!+ \left.\sum_{m=0}^{q-1} 2(m-q)\left(n(2q+1)+2m+1\right)\left[\left(n(2q+1)+2m+1\right)^2 +\left(\frac{2la\pi}{\beta}\right)^2\right]^{-s}\right\}\,.\label{d2q+1}\eeq

After changing $q-m$ by $m$ in the second term, and extracting some factors $2q+1$, we get
\beq
\delta\zeta_{S^3/{\mathbb Z}_{2q+1}}(s)&=&-2{\left(\frac{\mu\,a}{2q+1}\right)}^{2s}\sum_{\substack{l=-\infty\\ n=0}}^{\infty}\left\{\sum_{m=1}^q m\left(n+\frac{2m}{2q+1}\right)\left[\left(n+\frac{2m}{2q+1}\right)^2 +\left(\frac{2\pi a l}{(2q+1)\beta}\right)^2\right]^{-s}\right. \nn \\  &-& \left.\sum_{m=1}^{q} m\left(n+1-\frac{2m}{2q+1}\right)\left[\left(n+1-\frac{2m}{2q+1}\right)^2 +\left(\frac{2\pi a l}{(2q+1)\beta}\right)^2\right]^{-s}\right\}\,.\label{delta2q+1}\eeq

 We evaluate, in the first place, the contribution of the $l=0$ term. As we have already commented, this is the part of the effective action which, added to the corresponding term in the effective action of $S^3$ divided by $2q+1$, will give rise to the topological  subleading entropy. We have
 \beq
\!\!\!\!\!\!\!\!\!\!\!\!\!\!\delta\zeta_{S^3/{\mathbb Z}_{2q+1}}^{l=0}(s)&=& -2{\left(\frac{\mu\,a}{2q+1}\right)}^{2s}\sum_{m=1}^{q}m\left\{{\zeta}_H\left(2s-1,\frac{2m}{2q+1}\right)-{\zeta}_H\left(2s-1,1-\frac{2m}{2q+1}\right)\right\}  \label{deltazetaimpar}\,, \eeq
where ${\zeta}_H(z,q)$ is the Hurwitz zeta function.

That this piece of the relevant zeta function vanishes at $s=0$ can be easily shown by using the series representation \cite{rusa}, formula $9.521.2$ for the Hurwitz zeta, which, taking into account that $\cos{(s-\frac12)\pi}=\frac{\pi}{\Gamma(s)\Gamma(1-s)}$, gives as a result
 \beq
\delta\zeta_{S^3/{\mathbb Z}_{2q+1}}^{l=0}(s)=-8{\left(\frac{\mu\,a}{2q+1}\right)}^{2s}\frac{\Gamma(2-2s)({2\pi})^{2s-2}\pi\,s}{\Gamma(1+s)\Gamma(1-s)}\sum_{m=1}^{q}m\sum_{n=1}^{\infty}n^{2s-2}\,\sin{\left(\frac{4\pi nm}{2q+1}\right)}\,.\label{delta02q+1}\eeq

An alternative expression for \eqref{deltazetaimpar} is given by
 \beq\nn
\delta\zeta_{S^3/{\mathbb Z}_{2q+1}}^{l=0}(s)&=& -2{\left(\frac{\mu\,a}{2q+1}\right)}^{2s}\left\{\frac12\sum_{l=1}^{2q}l\,{\zeta}_H\left(2s-1,\frac{l}{2q+1}\right)-\frac{2q+1}{2}\sum_{m=0}^{q-1}{\zeta}_H\left(2s-1,\frac{2m+1}{2q+1}\right)\right\}  \,. \eeq
Such equivalence is easily shown by renaming indices and reordering terms.

The contribution from $l\neq 0$ terms in \eqref{delta2q+1} can be rewritten as
 \beq
\!\!\!\!\!&&\!\!\!\!\delta\zeta_{S^3/{\mathbb Z}_{2q+1}}^{l\neq 0}(s)=\nonumber\\
&=& -4{\left(\frac{\mu\,a}{2q+1}\right)}^{2s}\sum_{m=0}^{q-1} (m+1)\sum_{\substack{l=1\\ n=-\infty}}^{\infty}\!\!\!\!\left(n+\frac{2m+2}{2q+1}\right)\left\{\left(n+\frac{2m+2}{2q+1}\right)^2 +\left(\frac{2\pi a l}{(2q+1)\beta}\right)^2\right\}^{-s}\nn\\
                                                 &=& -\frac{{(\mu\,a)}^{2s}{(2q+1)}^{1-2s}}{(1-s)}\sum_{m=0}^{q-1} (m+1)\, \frac{d}{d\alpha}\left.\sum_{\substack{l=1\\ n=-\infty}}^{\infty}\left\{\left(n+\frac{2m+2\alpha}{2q+1}\right)^2 +\left(\frac{2\pi a l}{(2q+1)\beta}\right)^2\right\}^{-s+1}\right\vert_{\alpha=1}.\nn\eeq

Now, the double infinite sum can be extended analytically as done for similar expressions in the $S^3$ case, to obtain
 \beq
\!\!\!\!\delta\zeta_{S^3/{\mathbb Z}_{2q+1}}^{l\neq 0}\!\!(s)\!=\!\!\frac{-16{(\mu\,a)}^{2s}{\pi}^\frac32 s}{\Gamma(s\!+\!1){(2q\!+\!1)}^{2s}}\!\!\sum_{m=1}^{q} \!m\!\!\sum_{l,n=1}^{\infty}\!n\sin{\left(\frac{4\pi n m}{2q\!+\!1}\right)}\!\!\left(\frac{(2q\!+\!1)\beta n}{2al}\right)^{s-\frac32}
\!\!\!\!\!\!K_{s-\frac32}\!\!\left(\!\frac{ 4{\pi}^2aln}{ (2q\!+\!1)\beta}\!\right)\,.\label{delta12q+1}\eeq
Note that also this part of the zeta function vanishes at $s=0$.

Now, we can add the contributions to the effective action coming from equations \eqref{delta02q+1} and \eqref{delta12q+1}. We obtain
\beq
 \delta S_{\mathrm{eff},\,S^3/{\mathbb Z}_{2q+1}}&=&\frac{1}{\pi}\sum_{m=1}^{q}m\sum_{n=1}^{\infty}n^{-2}\,\sin{\left(\frac{4\pi nm}{2q+1}\right)}\nn \\
 &+&\frac{2}{\pi}\sum_{m=1}^{q}m\sum_{n=1}^{\infty}n^{-2}\,\sin{\left(\frac{4\pi nm}{2q+1}\right)}\sum_{l=1}^{\infty}\left(1+\frac{4\pi^2 aln}{(2q+1)\beta}\right)e^{-\frac{4\pi^2 aln}{(2q+1)\beta}}\,.\eeq

Note that the first term is the one remaining in the high-temperature limit, while the others are exponentially decreasing corrections. According to equation \eqref{2q+1}, we must add this to the effective action of the 3-sphere (equation \eqref{eff3}) divided by $2q+1$ to get the complete effective action for $S^3/{\mathbb Z}_{2q+1}$. Thus, we obtain

\beq
  S_{\mathrm{eff},\,S^3/{\mathbb Z}_{2q+1}}&=&\frac{1}{4{\pi}^2(2q+1)}{\zeta}_R (3)+
 \frac{1}{\pi}\sum_{m=1}^{q}m\sum_{n=1}^{\infty}n^{-2}\,\sin{\left(\frac{4\pi nm}{2q+1}\right)} -\frac{8{\pi}^4}{3(2q+1)} \left(\frac{a}{\beta}\right)^3{\zeta}_R (-3)\nn\\
 &+& \frac{1}{4\pi^2(2q+1)}\sum_{k,l=1}^{\infty}\frac{1}{k^3} \left[ 2 + 2\frac{4\pi^2akl}{\beta} + \left( \frac{4\pi^2akl}{\beta} \right)^2 \right] e^{-\frac{4\pi^2akl}{\beta}}\nn \\
 &+&\frac{2}{\pi}\sum_{m=1}^{q}m\sum_{n=1}^{\infty}n^{-2}\,\sin{\left(\frac{4\pi nm}{2q+1}\right)}\sum_{l=1}^{\infty}\left(1+\frac{4\pi^2 aln}{(2q+1)\beta}\right)e^{-\frac{4\pi^2 aln}{(2q+1)\beta}}\,.\label{eff2q+1}\eeq

In particular, in the $\beta \rightarrow 0$ limit one has
\beq S_{\mathrm{eff},\,S^3/{\mathbb Z}_{2q+1}}=\frac{{\zeta}_R (3)}{4{\pi}^2(2q+1)}\!+\!\frac{1}{\pi}\!\sum_{m=1}^{q}\!m\!\sum_{n=1}^{\infty}n^{-2}\sin{\left(\!\frac{4\pi nm}{2q+1}\!\right)}\!-\!\frac{8{\pi}^4{\zeta}_R (-3)}{3(2q+1)} \left(\!\frac{a}{\beta}\!\right)^3\!\! {+ {\cal O}{(e^{-\frac{a}{\beta}})}.}\label{eff2q+10}\eeq

The first and second terms are the ones contributing to the topological  subleading entropy, while the last term leads to a modified extensive entropy of Stefan-Boltzmann type.

Again the first two terms  of the effective action \eqref{eff2q+1}
come from  the determinant of the corresponding spatial operator.

\paragraph{Even lens spaces  $M=L(2q,1)$.}
\label{ss24}

In this case, the degeneracy of the eigenvalues $\lambda_k^2$ vanishes for $k$ even. Thus, it is very easy to relate the effective action for these homogeneous lens spaces to the one of $S^3$.

\bigskip

Let us first analyze the simplest case, $L(2,1)=S^3 /{\mathbb Z}_{2}$.
One has
\begin{equation}
 \zeta_{S^3/{\mathbb Z}_2}(s)= ({\mu\,a})^{2s} \sum_{l=-\infty}^\infty\sum_{n=0}^{\infty} (2n+1)^2 \left(\left(\frac{2\pi a l}{\beta}\right)^2+(2n+1)^2 \right)^{-s}\,.
 \end{equation}

Now, we can add and subtract to the previous expression the missing contribution from the even modes to get
\beq
 \zeta_{S^3/{\mathbb Z}_2}(s)= ({\mu\,a})^{2s} \sum_{l=-\infty}^\infty\sum_{k=1}^{\infty}\left\{k^2 \left(\left(\frac{2\pi a l}{\beta}\right)^2+k^2 \right)^{-s} -
 (2k)^2 \left(\left(\frac{2\pi a l}{\beta}\right)^2+(2k)^2 \right)^{-s}\right\}\,.
 \label{zetasz2}
 \eeq

 It is then clear that the first term between curly brackets is nothing but $ \zeta_{S^3}(s,\beta)$, while the second term, with opposite sign is $2^{2-2s} \zeta_{S^3}(s,2\beta)$.

 Taking into account that $\zeta_{S^3}(0)=0$ no matter the value of $\beta$, the effective action for this very simple lens space can be obtained from equation \eqref{eff3} as
 \beq
 S_{\mathrm{eff},\,S^3/{\mathbb Z}_2}=S_{\mathrm{eff},\,S^3}(\beta)-4S_{\mathrm{eff},\,S^3}(2\beta)\nn\,,
 \eeq
 which gives as a result
\beq
 S_{\mathrm{eff},\,S^3/{\mathbb Z}_2}&=&-\frac{3}{4{\pi}^2}{\zeta}_R (3)-\frac{4{\pi}^4}{3} \left(\frac{a}{\beta}\right)^3{\zeta}_R (-3)+\frac{1}{4\pi^2}\sum_{k,l=1}^{\infty}\frac{1}{k^3} \left[ 2 + 2\frac{4\pi^2akl}{\beta} + \left( \frac{4\pi^2akl}{\beta} \right)^2 \right] e^{-\frac{4\pi^2akl}{\beta}}\nn \\&-&\frac{1}{\pi^2}\sum_{k,l=1}^{\infty}\frac{1}{k^3} \left[ 2 + 2\frac{2\pi^2akl}{\beta} + \left( \frac{2\pi^2akl}{\beta} \right)^2 \right] e^{-\frac{2\pi^2akl}{\beta}}\,.
\label{effz2}
\eeq

As before, the high-temperature limit leaves a topological term in addition to the expected Stefan-Boltzmann one, and the corrections are exponentially decreasing.

\bigskip

The complete effective action for generic even lens spaces $L(2q,1)$ can be simplified by using the one of $L(2,1)$. In fact, the degeneracies are given by equation \eqref{even}. Using a new index $k=0,1,...,\infty$ to label the odd eigenvalues, we have
\[d_k=\left(2\left[\frac{2k+1}{2q}\right]+1\right)(2k+1)\,.\]
Now, we can write $k=nq+m$ with $n=0,1,...,\infty$ and $m=0,1,...,q-1$. Then,

\[d_k=\left(2\frac{k-m}{q}+1\right)(2k+1)=\frac{(2k+1)^2}{q}+\left(\frac{q-m-1}{q}-\frac{m}{q}\right)(2k+1)\,.\]

The first term is nothing but the degeneracy in the case $S^3 /{\mathbb Z}_{2}$, divided by $q$. Thus, the total zeta function is given by
\beq\zeta_{S^3/{\mathbb Z}_{2q}}(s)=\frac{\zeta_{S^3 /{\mathbb Z}_{2}}(s)}{q}+\delta\zeta_{S^3/{\mathbb Z}_{2q}}(s)\,,\label{2q}\eeq
with
\beq
\!\!\!\!\!\delta\zeta_{S^3/{\mathbb Z}_{2q}}\!(s)=2{\left(\frac{\mu\,a}{2q}\right)}^{2s}\!\!\!\sum_{\substack{l=-\infty\\n=0}}^{\infty}\sum_{m=0}^{q-1} \!\left((q-m-1)-m\right)\!\left(n+\frac{2m+1}{2q}\right)\!\!\left[\!\left(\!n\!+\!\frac{2m+1}{2q}\!\right)^2 \!+\!\left(\!\frac{\pi al}{q\beta}\!\right)^2\right]^{-s} \label{d2q}  \eeq
 or, changing $m\rightarrow q-m-1$ in the first term,
 \beq
\delta\zeta_{S^3/{\mathbb Z}_{2q}}(s)=2{\left(\frac{\mu\,a}{2q}\right)}^{2s}\sum_{\substack{l=-\infty\\n=0}}^{\infty}\sum_{m=1}^{q-1} m\left\{\left(n+1-\frac{2m+1}{2q}\right)\left[\left(n+1-\frac{2m+1}{2q}\right)^2 +\left(\frac{\pi al}{q\beta}\right)^2\right]^{-s}\right.\nn\\-\left.\left(n+\frac{2m+1}{2q}\right)\left[\left(n+\frac{2m+1}{2q}\right)^2 +\left(\frac{\pi al}{q\beta}\right)^2\right]^{-s} \right\}\label{delta2q}\nn \,.\eeq

As done in the previous cases, we evaluate first the contribution of the $l=0$ term which, added to the corresponding term in the effective action of $S^3/{\mathbb Z}_2$ divided by $q$, will give rise to the topological  subleading entropy. We have

 \beq
\!\!\!\delta\zeta_{S^3/{\mathbb Z}_{2q}}^{l=0}(s)&=& -2{\left(\frac{\mu\,a}{2q}\right)}^{2s}\sum_{m=1}^{q-1}m\left\{{\zeta}_H\left(2s-1,\frac{2m+1}{2q}\right)-{\zeta}_H\left(2s-1,1-\frac{2m+1}{2q}\right)\right\} \label{deltazetapar} \,. \eeq

Again, this piece of the relevant zeta function vanishes at $s=0$ and can be written as
 \beq
\delta\zeta_{S^3/{\mathbb Z}_{2q}}^{l=0}(s)=-8{\left(\frac{\mu\,a}{2q}\right)}^{2s}\frac{\Gamma(2-2s)\pi\,s}{({2\pi})^{2-2s}\Gamma(1+s)\Gamma(1-s)}\sum_{m=1}^{q-1}m\sum_{n=1}^{\infty}n^{2s-2}\,\sin{\frac{2\pi n(2m+1)}{2q}}\,.\label{delta02q}\eeq

An alternative expression for \eqref{deltazetapar} can be obtained by renaming indices, reordering terms and, finally, using a multiplication theorem for Hurwitz zeta function (see \cite{Abramowitz,nistdl} and references therein),

 \beq
\delta\zeta_{S^3/{\mathbb Z}_{2q}}^{l=0}(s)&=& -2{\left(\frac{\mu\,a}{2q}\right)}^{2s}\left\{\sum_{m=0}^{q-1}(2m+1){\zeta}_H\left(2s-1,\frac{2m+1}{2q}\right)-q^{2s}{\zeta}_H\left(2s-1,\frac12\right)\right\} 
\,, \nn \eeq
which added to the corresponding contribution of $S^3/\mathbb{Z}_2$ reproduces the result obtained for the zeta function of the spatial operator in \cite{Dowker041}.

The contribution from $l\neq 0$, evaluated with the same techniques used in the odd case, gives as a result
 \beq
\!\!\!\delta\zeta_{S^3/{\mathbb Z}_{2q}}^{l\neq 0}\!(s)\!=\!\frac{-16\pi^{3/2}s}{q\Gamma(s\!+\!1)} {\left(\frac{\mu a}{2q}\right)}^{2s}
 \sum_{m=1}^{q-1}\!m\!\sum_{l,n=1}^{\infty}\!\!n\sin{\left(2\pi n\frac{2m\!+\!1}{2q}\right)}
 \left(\!\frac{q\beta n}{al}\!\right)^{s-\frac32}\!\!\!K_{s-\frac{3}{2}}\left(\!\frac{2\pi^2aln}{q\beta}\!\right).\label{delta12q}\eeq

Now, we can determine the contribution to the effective action coming from equations \eqref{delta02q} and \eqref{delta12q}, which is
\beq
 \delta S_{\mathrm{eff},\,S^3/{\mathbb Z}_{2q}}&=&\frac{1}{\pi}\sum_{m=1}^{q-1}m\sum_{n=1}^{\infty}n^{-2}\,\sin{\left(2\pi n\frac{2m+1}{2q}\right)}\nn \\
 &+&\frac{2}{\pi q}\sum_{m=1}^{q-1}m\sum_{n,l=1}^{\infty}n^{-2}\,\sin{\left(2\pi n\frac{2m+1}{2q}\right)}\left(1+\frac{2{\pi}^2aln}{\beta q}\right)e^{-\frac{2{\pi}^2aln}{\beta q}}\,.\eeq

 Also in this case, the first term is the one remaining in the high-temperature limit, and the others are exponentially decreasing corrections. According to equation \eqref{2q}, we must add this to the effective action on $S^3/{\mathbb Z}_2$ (equation \eqref{effz2}) divided by $q$ to get the complete effective action on $S^3/{\mathbb Z}_{2q}$. Thus, we get

\beq
S_{\mathrm{eff},\,S^3/{\mathbb Z}_{2q}}&=&
-\frac{3}{4q{\pi}^2}{\zeta}_R (3)-\frac{4{\pi}^4}{3q} \left(\frac{a}{\beta}\right)^3{\zeta}_R (-3) +\frac{1}{\pi}\sum_{m=1}^{q-1}m\sum_{n=1}^{\infty}n^{-2}\,\sin{\left(2\pi n\frac{2m+1}{2q}\right)} \nn \\ &+&\frac{1}{4\pi^2 q}\sum_{k,l=1}^{\infty}\frac{1}{k^3} \left[ 2 + 2\frac{4\pi^2akl}{\beta} + \left( \frac{4\pi^2akl}{\beta} \right)^2 \right] e^{-\frac{4\pi^2akl}{\beta}}\nn \\&-&\frac{1}{q\pi^2}\sum_{k,l=1}^{\infty}\frac{1}{k^3} \left[ 2 + 2\frac{2\pi^2akl}{\beta} + \left( \frac{2\pi^2akl}{\beta} \right)^2 \right] e^{-\frac{2\pi^2akl}{\beta}}\nn\\
&+&\frac{2}{\pi q}\sum_{m=1}^{q-1}m\sum_{n,l=1}^{\infty}n^{-2}\,\sin{\left(2\pi n\frac{2m+1}{2q}\right)}\left(1+\frac{2{\pi}^2aln}{\beta q}\right)e^{-\frac{2{\pi}^2aln}{\beta q}}.\label{eff2q}\eeq

In particular, in the $\beta \rightarrow 0$ limit, one has
\beq
S_{\mathrm{eff},\,S^3/{\mathbb Z}_{2q}}\!=\!-\frac{3{\zeta}_R (3)}{4{\pi}^2q}\!+\!\frac{1}{\pi}\!\sum_{m=1}^{q-1}\!m\!\sum_{n=1}^{\infty}n^{-2}\sin{\left(\!2\pi n\frac{2m+1}{2q}\!\right)}\!-\!\frac{4{\pi}^4{\zeta}_R (-3)}{3q} \left(\!\frac{a}{\beta}\!\right)^3\!\!\!{+{\cal O}{(e^{-\frac{a}{\beta}})}.}\eeq

The first and second terms are the ones contributing to the topological  subleading entropy, while the last term leads to a modified extensive entropy.

Again the first two terms are obtained from the determinant of the spatial operator, and are equivalent to the result in \cite{Dowker041}.

\section{An alternative expression for $S_{\mathrm{eff}}$ and the low-temperature limit}
\label{low}

As already said, in this section we obtain alternative expressions for the effective actions, which are adequate for taking the low-temperature limit.
This time we use the Jacobi-Poisson formula to invert the infinite sums over $l$ instead of those over $k$. Let us begin by analyzing the field theory on the 3-sphere.

\paragraph{3-sphere $M=S^3$.}

We rewrite the relevant zeta function as
\beq
 \zeta_{S^3}(s) = \left(\frac{2\pi}{\mu\beta}\right)^{-2s}\sum_{l=-\infty}^{\infty}\sum_{k=1}^{\infty} k^2 \left[
 \left(\frac{k\beta}{2\pi a}\right)^2 + l^2 \right]^{-s}
\nn\eeq
or, using the Mellin transform,
\beq\label{s}
  \zeta_{S^3}(s) = \frac{1}{\Gamma(s)} \left(\frac{2\pi}{\mu\beta}\right)^{-2s}\sum_{l=-\infty}^{\infty}\sum_{k=1}^{\infty}
  k^2 \int_0^{\infty} dt\, t^{s-1} e^{-\left[
 \left(\frac{k\beta}{2\pi a}\right)^2 + l^2 \right]t}\,.
\nn\eeq

Here, a direct application of \eqref{kk} leads to

\beq\nonumber
 \zeta_{S^3}(s) = \frac{\pi^{1/2}}{\Gamma(s)} \left(\frac{2\pi}{\mu\beta}\right)^{-2s} \sum_{k=1}^{\infty} k^2 \left\{\int_0^{\infty} dt\, t^{s-\frac{3}{2}}
 e^{-\left(\frac{k\beta}{2\pi a}\right)^2t} + 2 \sum_{l=1}^{\infty} \int_0^{\infty} dt\, t^{s-\frac{3}{2}} e^{-\left(\frac{k\beta}{2\pi a}\right)^2t}
 e^{-\frac{\pi^2l^2}{t}}\right\}\,,\nn
\eeq
where we made the term corresponding to $l=0$ explicit. This is the one which will remain in the zero-temperature limit. Once the integral is performed, it reads
\beq
 \zeta_{S^3}^{l=0}(s) = s \frac{\beta\mu^{2s}}{2\pi^{1/2}} \frac{\Gamma(s-\frac{1}{2})}{\Gamma(s+1)} \sum_{k=1}^{\infty}\frac{k^{3-2s}}{a^{1-2s}}\,.\nn
\eeq

For the remaining terms the integral can also be performed, to obtain
\beq
 \zeta_{S^3}^{l\neq 0}(s) =
 s\frac{2\pi^{1/2}}{\Gamma(s+1)}\left(\frac{2\pi}{\mu\beta}\right)^{-2s}\sum_{k=1}^{\infty}k^2 \sum_{l=1}^{\infty} 2
 \left(\frac{2\pi^2al}{k\beta}\right)^{s-\frac{1}{2}}K_{s-\frac{1}{2}}\left(\tfrac{k\beta l}{a}\right).
\eeq

Now, performing the $s$-derivative, the total effective action can be expressed as
\beq
 S_{\textrm{eff},\,S^3} = \frac{\beta}{2a}\zeta_R(-3) -2\pi^{1/2}\sum_{k=1}^{\infty} k^2 \sum_{l=1}^{\infty}
 \left(\frac{2\pi^2al}{k\beta}\right)^{-1/2} K_{\frac{1}{2}}\left(\tfrac{k\beta l}{a}\right)\nn
\eeq
or, using the explicit expression of the modified Bessel function,
\beq
S_{\textrm{eff},\,S^3} = \frac{\beta}{240 a} -\sum_{k=1}^{\infty} k^2 \sum_{l=1}^{\infty}
 l^{-1}e^{-\frac{\beta kl}{a}}\,.\label{low3}
\eeq

Finally, the sum over $l$ can be performed, giving as a result
\beq
S_{\textrm{eff},\,S^3} = \frac{\beta}{240a} + \sum_{k=1}^{\infty} k^2 \log\left(1-e^{-k\beta/a}\right)\label{eff3low}.
\eeq

In the appendix \ref{ap1} it is shown  how, from this expression, it is possible to recover the high-temperature expansion by means of Abel-Plana formula.

\bigskip

The last form of the effective action was to be expected since, for any compact manifold, one has
\beq
\label{finite}
S_{\rm eff}=\beta E_0+\sum_{k=1}^\infty d_k \log(1-e^{-\beta \lambda_k})\,,
\eeq
where $\lambda_k^2$ are the eigenvalues of minus the conformal Laplacian, $d_k$ are their degeneracies and $E_0$ is the regularized vacuum energy at zero temperature.
Let us  show that the same expansion can be derived for any spherical space $S^3/\Gamma^\ast$ in the effective
action approach, with  $\Gamma^*$ any discrete subgroup of $SU(2)$. This includes the cyclic  groups ${\mathbb Z}_q$
but also the binary cubic ${\mathbb D}^\ast_q$, tetrahedral ${\mathbb T}^\ast$, octahedral ${\mathbb O}^\ast$, and icosahedral  groups ${\mathbb Y}^\ast$ \cite{acm}.
In those cases $\lambda_k^2$ are the corresponding eigenvalues of
the conformal laplacian $-\Delta + \frac1{a^2}$ acting on the space of square integrable functions on $S^3/\Gamma^\ast$ and
$d_k$ their degeneracies.

We rewrite the relevant zeta function as
\beq
 \zeta_{S^3/\Gamma^\ast}(s) = \left(\frac{2\pi}{\mu\beta}\right)^{-2s}\sum_{l=-\infty}^{\infty}\sum_{k=1}^{\infty} d_k \left[
 \left(\frac{\lambda_k\beta}{2\pi}\right)^2 + l^2 \right]^{-s}
\nn\eeq
or, using the Mellin transform,
\beq\label{sgen}
  \zeta_{S^3/\Gamma^\ast}(s) = \frac{1}{\Gamma(s)} \left(\frac{2\pi}{\mu\beta}\right)^{-2s}\sum_{l=-\infty}^{\infty}\sum_{k=1}^{\infty}
  d_k \int_0^{\infty} dt\, t^{s-1} e^{-\left[
 \left(\frac{\lambda_k\beta}{2\pi}\right)^2 + l^2 \right]t}\,.
\nn\eeq

A direct application of \eqref{kk} leads to

\beq\nonumber
 \zeta_{S^3/\Gamma^\ast}(s) = \frac{\pi^{1/2}}{\Gamma(s)} \left(\frac{2\pi}{\mu\beta}\right)^{-2s} \sum_{k=1}^{\infty} d_k \left\{\int_0^{\infty} dt\, t^{s-\frac{3}{2}}
 e^{-\left(\frac{{\lambda_k}\beta}{2\pi}\right)^2t} + 2 \sum_{l=1}^{\infty} \int_0^{\infty} dt\, t^{s-\frac{3}{2}} e^{-\left(\frac{{\lambda_k}\beta}{2\pi}\right)^2t}
 e^{-\frac{\pi^2l^2}{t}}\right\}\,.\nn
\eeq
Here, we made the $l=0$ term explicit. This is the one which will remain in the zero temperature limit. Once the integral is performed, it reads
\beq
 \zeta_{S^3/\Gamma^\ast}^{l=0}(s) = s \frac{\beta\mu^{2s}}{2\pi^{1/2}} \frac{\Gamma(s-\frac{1}{2})}{\Gamma(s+1)}\sum_{k=1}^{\infty} d_k {\lambda_k}^{1-2s}\,.\nn
\eeq
For the remaining terms, the integral can also be performed, to obtain
\beq
 \zeta_{S^3/\Gamma^\ast}^{l\neq 0}(s) =
 s\frac{2\pi^{1/2}}{\Gamma(s+1)}\left(\frac{2\pi}{\mu\beta}\right)^{-2s}\sum_{k=0}^{\infty}d_k \sum_{l=1}^{\infty} 2
 \left(\frac{2\pi^2l}{{\lambda_k}\beta}\right)^{s-\frac{1}{2}}K_{s-\frac{1}{2}}\left({\lambda_k}\beta l\right).
\eeq

Now, performing the $s$-derivative, the total effective action can be expressed as
\beq
 S_{\mathrm{eff},\,S^3/\Gamma^\ast} =
 \beta E_c -2\pi^{1/2}\sum_{k=1}^{\infty} d_k \sum_{l=1}^{\infty}
 \left(\frac{2\pi^2l}{{\lambda_k}\beta}\right)^{-1/2} K_{\frac{1}{2}}\left({\lambda_k}\beta l\right),\nn
\eeq
where $E_c$ is the Casimir energy, or, using the explicit expression of the modified Bessel function,
\beq
S_{\mathrm{eff},\,S^3/\Gamma^\ast} = \beta  E_c -\sum_{k=1}^{\infty} d_k \sum_{l=1}^{\infty}
 l^{-1}e^{-{\lambda_k}\beta l}\,.\label{low3gen}
\eeq

Now, the sum over $l$ can be performed, giving as a result
\beq
S_{\mathrm{eff},\,S^3/\Gamma^\ast} = \beta  E_c+ \sum_{k=1}^{\infty} d_k \log\left(1-e^{-{\lambda_k}\beta}\right).
\label{fintem}
\eeq

As we will see later, in  the case of lens spaces $L(p,1)$ the value of the Casimir energy
can be easily evaluated \cite{Dowker78,Dowker89,Dowker041}
\beq
E_c= - \frac{p^4 + 10 p^2 -14}{
720 p} \frac1{a}. \label{edowker}\eeq
The remaining terms are exponentially suppressed, as can be seen from \eqref{fintem}.

\paragraph{Odd lens spaces $M=L(2q+1,1)$.}
\label{ss32}

As shown in the previous section
\beq\zeta_{S^3/{\mathbb Z}_{2q+1}}(s)=\frac{\zeta_{S^3}(s)}{2q+1}+\delta\zeta_{S^3/{\mathbb Z}_{2q+1}}(s)\,.\eeq
where, starting from \eqref{d2q+1}, replacing $m$ with $q-m$ in the second term and extracting a convenient overall factor, we have
\beq\delta\zeta_{S^3/{\mathbb Z}_{2q+1}}\!(s)\!&=\!&
\frac{-2({\mu\,\beta})^{2s}}{(2q\!+\!1){(2\pi )}^{2s}}\!\sum_{\substack{\l=-\infty\\n=0}}^{\infty}
                                                 \!\left\{\sum_{m=1}^q\!m\left(n(2q\!+\!1)+2m\right)\left[\left(\frac{\beta}{2a\pi}\right)^2\left(n(2q\!+\!1)+2m\right)^2 +l^2\right]^{-s}\right. \nn \\  &-& \left.\sum_{m=1}^{q} m\left((n+1)(2q+1)-2m\right)\left[\left(\frac{\beta}{2a\pi}\right)^2\left((n+1)(2q+1)-2m\right)^2 +l^2\right]^{-s}\right\}\,.
                                                 \label{you}
                                                 \eeq

The infinite sum over $l$ can be inverted using, again, equation \eqref{kk}, and we have
 \beq\delta\zeta_{S^3/{\mathbb Z}_{2q+1}}(s)&=&-\frac{2 (\mu\, \beta)^{2s}}{(2q+1){({2\pi })^{2s}}}\frac{s{\pi}^{\frac12}}{\Gamma(s+1)}\sum_{m=1}^q m \sum_{l=-\infty}^{\infty}
                                            \sum_
                                                  {n=0}^{\infty}
                                                  \\                                               & &
                                         \left\{\left[n(2q+1)+2m\right]   \int_{0}^{\infty}dt\,t^{s-\frac32}\,e^{-\frac{l^2 \pi^2}{t}}    e^{-t\left(\frac{\beta}{2a\pi}\right)^2\left[n(2q+1)+2m\right]^2}\right. \nn \\  &-& \left. \left[(n+1)(2q+1)-2m\right]\int_{0}^{\infty}dt\,t^{s-\frac32}\,e^{-\frac{l^2 \pi^2}{t}}e^{-t\left(\frac{\beta}{2a\pi}\right)^2\left[(n+1)(2q+1)-2m\right]^2}\right\}\nn\,.\eeq

 The evaluation of the effective action is, from this point on, entirely analogous to the ones already performed. Thus, we merely give the final result for $\delta S_{\mathrm{eff},\,S^3/{\mathbb Z}_{2q+1}}$, which is
 \beq
 & &\delta S_{\mathrm{eff},\,S^3/{\mathbb Z}_{2q+1}}=\frac{(2q+1)\beta}{a\pi^3}\sum_{m=1}^q m \sum_{n=1}^{\infty}n^{-3}\sin{\left(\frac{4\pi nm}{2q+1}\right)}\nn\\&+&2\sum_{m=1}^q m \sum_{\substack{l=1\\n=0}}^{\infty}\frac{1}{l}\left\{\left(n+\frac{2m}{2q+1}\right)e^{-\frac{\beta l(2q+1)}{a}\left(n+\frac{2m}{2q+1}\right)}-\left(n+1-\frac{2m}{2q+1}\right)e^{-\frac{\beta l(2q+1)}{a}\left(n+1-\frac{2m}{2q+1}\right)}\right\}\nn\eeq

 As usual, the first term, which is the contribution from $l=0$, is the one which will remain (together with the corresponding one in the effective action of $S^3$ divided by $2q+1$) in the zero temperature limit. The remaining terms are exponential corrections.

\bigskip

 Thus, the complete effective action at low temperatures is given by
 \beq S_{\mathrm{eff},\,S^3/{\mathbb Z}_{2q+1}}&=&\frac{\beta}{2a(2q+1)}\zeta_R(-3) +\frac{(2q+1)\beta}{a\pi^3}\sum_{m=1}^q m \sum_{n=1}^{\infty}n^{-3}\sin{\left(\frac{2\pi nm}{q+\frac12}\right)}{+{\cal O}{(e^{-\frac{\beta}{a}})}}\nn \\
&=&\frac{\beta}{2a(2q+1)}\zeta_R(-3)-\frac{3 q + 4 q^2 + 2 q^3 + q^4}{45(2q +1)}\frac{\beta}{a}\, {+{\cal O}{(e^{-\frac{\beta}{a}})}},\nn
\eeq
where the second line is easily obtained performing first the sum over $n$ in the first identity.
The second term is what we will call the `genuine topological contribution' in the sense that it comes from non-trivial holonomies. This equation can also be written as
\beq
S_{\mathrm{eff},\,S^3/{\mathbb Z}_{2q+1}}&=&\frac{14 - 10 (1 + 2 q)^2 - (1 + 2 q)^4}{720 (1 + 2 q)}\frac{\beta}{a} \,{ +\, {\cal O}{(e^{-\frac{\beta}{a}})}},\nn
\eeq
which is a particular case of the result \eqref{edowker}. Here, it is clear that the first term, which depends on $1/(2q+1)$, is the only one that can be interpreted as the integral of a local density \cite{bd,acm}.

\paragraph{Even lens spaces $M=L(2q,1)$.}
\label{ss34}

Let us first evaluate the  $L(2,1)$ case.
As done for high temperatures, we make use of

\beq S_{\mathrm{eff},\,S^3/{\mathbb Z}_2}=S_{\mathrm{eff},\,S^3}(\beta)-4S_{\mathrm{eff},\,S^3}(2\beta)\nn\eeq
to obtain

\beq
S_{\mathrm{eff},\,S^3/{\mathbb Z}_2} = -\frac{7\beta}{2a}\zeta_R(-3) + \sum_{k=1}^{\infty} k^2 \log\left(1-e^{-k\beta/a}\right)-4 \sum_{k=1}^{\infty} k^2 \log\left(1-e^{-2k\beta/a}\right)\,.
\eeq

\bigskip

For $L(2q,1)$, one has

\beq\zeta_{S^3/{\mathbb Z}_{2q}}(s)=\frac{\zeta_{S^3 /{\mathbb Z}_{2}}(s)}{q}+\delta\,\zeta_{S^3/{\mathbb Z}_{2q}}(s)\,.\label{2q1}\eeq

Starting from \eqref{d2q}, changing $m\rightarrow q-m-1$ in the first term and extracting convenient overall factors, we have
 \beq
\delta\zeta_{S^3/{\mathbb Z}_{2q}}(s)=2{\left(\frac{\mu\beta}{2\pi}\right)}^{2s}\sum_{\substack{l=-\infty\\n=0}}^{\infty}\sum_{m=0}^{q-1} m\left\{\left(n+1-\frac{2m+1}{2q}\right)\left[\left(\frac{q\beta}{\pi a}\right)^2\left(n+1-\frac{2m+1}{2q}\right)^2 +l^2\right]^{-s}\right.\nn\\-\left.\left(n+\frac{2m+1}{2q}\right)\left[\left(\frac{q\beta}{\pi a}\right)^2\left(n+\frac{2m+1}{2q}\right)^2 +l^2\right]^{-s} \right\}\label{delta2qlow}\nn \,.\eeq

The analytic extension of this piece of the effective action is almost identical to the one presented in the odd case, thus we will only give the final result,

\beq
 \delta S_{\mathrm{eff},\,S^3/{\mathbb Z}_{2q}}&=&\frac{2q\beta}{a\pi^3}\sum_{m=1}^{q-1} m \sum_{n=1}^{\infty}n^{-3}\sin{\left(2\pi n\frac{2m+1}{2q}\right)}\nn\\&-2&\sum_{m=1}^{q-1} m \sum_{\substack{l=1\\n=0}}^{\infty}\frac{1}{l}\left\{\left(n+\frac{2m+1}{2q}\right)e^{-\frac{2\beta ql}{a}\left(n+\frac{2m+1}{2q}\right)}-\left(n+1-\frac{2m+1}{2q}\right)e^{-\frac{2\beta ql}{a}\left(n+1-\frac{2m+1}{2q}\right)}\right\}\,.\nn\eeq

 Also here, the first term is the one remaining at zero temperature, while the others are exponential corrections. When the first contribution is added to the corresponding one coming from the effective action of $L(2,1)$ divided by $q$ one has, at low temperatures,
 \beq
 S_{\mathrm{eff},\,S^3/{\mathbb Z}_{2q}}&=& -\frac{7\beta}{2qa}\zeta_R(-3)+\frac{2q\beta}{a\pi^3}\sum_{m=1}^{q-1} m \sum_{n=1}^{\infty}n^{-3}\sin{\left(2\pi n\frac{2m+1}{2q}\right)} {+{\cal O}{(e^{-\frac{\beta}{a}})}}\nn\\
&=&\frac{ \beta}{4 qa}\zeta_R(-3)+\frac{11 - 40 q^2 - 16 q^4}{1440\, q}\frac{\beta}{a}{+{\cal O}{(e^{-\frac{\beta}{a}})}}.\nn
\eeq
Since we have explicitly separated the contribution coming from the sphere divided by $2q$, the second term is again the genuine topological contribution.
Note that this last equation can also be written as
\beq
S_{\mathrm{eff},\,S^3/{\mathbb Z}_{2q}}&=&\frac{14 - 40 q^2 - 16 q^4}{1440\, q}\frac{\beta}{a}{+{\cal O}{(e^{-\frac{\beta}{a}})}},\nn
\eeq
which once more is a particular case of \eqref{edowker}. As in the odd case, only the first term can be interpreted as the integral of a local density \cite{bd,acm}.

\section{Thermodynamic properties and thermal duality}
\label{thermo}

From the general expressions obtained in the preceding sections, the thermodynamic properties of homogeneous lens spaces can be studied at any temperature. In particular, we will evaluate the free energy $F=\frac{1}{\beta}S_{\mathrm{eff}}$, the internal energy $E=\partial_{\beta}S_{\mathrm{eff}}$ and the entropy $S=\beta (E-F)$ in all cases, both in the low- and in the high-temperature limit.

\bigskip

We start with $S^3$.
At low temperature we find
\beq
F=E=\frac{1}{240a}\qquad S=0, \qquad \beta\rightarrow \infty\,,\nn\eeq
where we have used that $\zeta(-3)=1/120$.

In the infinite temperature limit, we have
\beq
F=\frac{\zeta_R(3)}{4\pi^2 \beta}-\frac{\pi^4 a^3}{45\beta^4}\qquad E=\frac{\pi^4 a^3}{15\beta^4}\qquad S=-\frac{\zeta_R(3)}{4\pi^2}+\frac{4\pi^4 a^3}{45\beta^3}, \qquad \beta\rightarrow 0\,.\nn\eeq

The results at low temperatures are very well-known. In particular, the value of the internal (vacuum or Casimir) energy coincides with the ones obtained through other types of regularization. The entropy vanishes in this limit, in agreement with the first law of Thermodynamics. As already commented, in the infinite temperature limit, apart from the usual extensive or Stefan-Boltzmann entropy (which is positive), a volume independent, negative, entropy appears, which is missing, for instance, in references \cite{kutasov,emilio}. We call this term  topological  subleading  entropy.

A very interesting fact appears when defining a new variable $\xi=2\pi a/\beta$ and looking at $\tilde{E}(\xi)=(aE)(\xi)$. Then, one can verify that the values of the internal energy in low- and high-temperature limits satisfy the thermal duality  law ${\xi}^{-2}\tilde{E}(\xi)={\xi}^{2}\tilde{E}(1/\xi)$. This indeed holds for the 3-sphere also at arbitrary finite values of $\xi$, as can be easily verified by performing the necessary derivative in equations \eqref{eff3} and \eqref{eff3low}. That this ``conformal" invariance applies in the case of the 3-sphere was proved and discussed time ago in ref. \cite{dowker}.

\bigskip

Now we analyze $L(2q+1,1)$.
At low temperatures,
\beq
F=E=\frac{14 - 10 (1 + 2 q)^2 - (1 + 2 q)^4}{720 a (1 + 2 q)} \qquad S=0, \qquad \beta\rightarrow \infty\,.\nn
\eeq

In the high-temperature limit, we have
\beq
F&=&\frac{{\zeta}_R (3)}{4\beta (2q+1){\pi}^2}+\frac{1}{\pi \beta}\sum_{m=1}^{q}m\sum_{n=1}^{\infty}n^{-2}\,\sin{\left(\frac{4\pi\, n\,m}{2q+1}\right)}-\frac{{\pi}^4}{45(2q+1)} \frac{a^3}{\beta^4}\,,\nn \\E&=&\frac{\pi^4 a^3}{15(2q+1)\beta^4}\,,\nn \\
S&=&-\frac{{\zeta}_R (3)}{4(2q+1){\pi}^2}-\frac{1}{\pi }\sum_{m=1}^{q}m\sum_{n=1}^{\infty}n^{-2}\,\sin{\left(\frac{4\pi\, n\,m}{2q+1}\right)}+\frac{{4\pi}^4}{45(2q+1)} \left(\frac{a}{\beta}\right)^3, \qquad \beta\rightarrow 0\,.\eeq

This case also satisfies the first law of Thermodynamics. As in the case of the 3-sphere, a topological  subleading  entropy appears in the infinite temperature limit, together with a positive Stefan-Boltzmann contribution. But, unlike in the covering space $S^3$, the thermal duality law for the internal energy fails in Lens spaces $L(p,1)$. This can be checked by comparing the high and the low temperature limits, since the extra term in the Casimir energy does not appear with an adequate $\xi$-dependence in the infinite temperature limit.

\bigskip

Finally, we go to $L(2q,1)$.
At low temperatures,
\beq
F=E=
\frac{14 - 40 q^2 - 16 q^4}{1440\,a q}
\qquad S=0, \qquad \beta\rightarrow \infty\,,\nn\eeq
while in the high-temperature limit we have
\beq
F&=&-3\frac{{\zeta}_R (3)}{4q\beta {\pi}^2}-\frac{{\pi}^4a^3}{90q\beta^4}+\frac{1}{\pi\beta}\sum_{m=1}^{q-1}m\sum_{n=1}^{\infty}n^{-2}\,\sin{\left(2\pi n\frac{2m+1}{2q}\right)}\,,\nn \\E&=&\frac{\pi^4 a^3}{30q\beta^4}\,,\nn \\
S&=&\frac{3{\zeta}_R (3)}{4q{\pi}^2}+\frac{2\pi^4}{45q} \left(\frac{a}{\beta}\right)^3-\frac{1}{\pi}\sum_{m=1}^{q-1}m\sum_{n=1}^{\infty}n^{-2}\,\sin{\left(2\pi n\frac{2m+1}{2q}\right)}, \qquad \beta\rightarrow 0\,.\eeq
The thermodynamical quantities for $L(2,1)$ can be obtained from the last three expressions by taking into account that the sum over $m$ does not appear when $q=1$.

Once more, the entropy vanishes at zero temperature. In the infinite temperature limit, the Stefan-Boltzmann entropy is modified, as compared to the case of the 3-sphere, by the expected volume factor. A topological  subleading  entropy appears, as in the previous cases, and there is no invariance under temperature inversion.

{ Notice that although  the topological subleading entropy  
term is temperature independent, the total entropy  is always monotonically increasing with the temperature. 
The whole entropy is also monotonically increasing with the internal energy in the high-temperature limit  and its derivative w.r.t. the internal energy at constant volume is nothing but $\beta$.  The agreement with thermodynamic fundamental laws is complete. }


{


Another interesting property of the subleading topological entropy is its subbaditivity. It does not saturate in this case due to the term with the double-sum. Indeed, the sum of the subleading entropies of $p$ copies of the lens space $L(p,1)$ is always greater than the subleading entropy of $S^3$.
}

\section{Conclusions}

In this paper we have given two different analytic expressions for the effective action of a conformally coupled massless scalar field on homogeneous lens spaces, for the whole temperature range.

The high- (low-) temperature
expansions of the thermodynamical quantities show that the leading terms have a power-like
dependence on  ${1/\beta}$ (${\beta}$) and the remaining corrections
are exponentially suppressed.

In the high-temperature case the leading term is given by Stefan-Boltzmann law and its coefficient
is an extensive quantity which is proportional to the space volume. For the entropy, the next order correction is temperature- and volume-independent
and only depends on the space topology. This contribution does not appear in the internal energy and
defines the topological  subleading  entropy, which characterizes
the topology of the space. We have shown that it is directly given by the determinant of the spatial
operator $-\Delta+\frac{1}{a^2}$ on the lens space. In the case of the sphere $S^3$
the topological subleading entropy is also directly related with the extensive
entropy of the same field theory defined in one less dimension space, as can be seen in appendix \ref{ap1}.

In the low-temperature limit, the leading contribution is the Casimir energy. Only part of it can be interpreted as the integral of a local density. Moreover, we have identified in all cases a genuine topological contribution, which comes from the holonomies and thus uniquely characterizes the lens space, in agreement with the results of ref.
\cite{acm}.

The presence of this topological contributions to the free energy and the entropy
introduce a breaking of thermal duality unlike in conformal theories in 1+1 dimensions.
Only in a pure spherical background $S^3$ this thermal duality is preserved.
As in the $S^3$ case \cite{kutasov}, the Cardy-Verlinde formula fails in lens spaces.

The same behavior is expected for the same theory on other topologically non-trivial spherical
backgrounds obtained by quotient of $S^3$  by binary cubic ${\mathbb D}^\ast_q$, tetrahedral ${\mathbb T}^\ast$, octahedral ${\mathbb O}^\ast$, and icosahedral ${\mathbb Y}^\ast$ discrete subgroups of  $SO(3)$ \cite{acm}
which deserve further study \cite{abds}.

\acknowledgments{The authors thank In\'es Cavero, Klaus Kirsten and Jos\'e Mar\'{\i}a Mu\~noz-Casta\~neda for useful discussions. {They also acknowledge the referee for his deep remarks. In particular for pointing out the connection between the subleading topological entropy and  Stefan-Boltzmann  entropy density in one dimension less.}
Research of CGB, DD and EMS was partially supported by CONICET PIP1787, ANPCyT PICT909 and UNLP Proyecto 11/X615 (Argentina). The work of MA has been partially supported by the Spanish MICINN grants FPA2009-09638 and CPAN Consolider Project CDS2007-42 and DGA-FSE (grant 2011-E24/2).}

\appendix

\section{Effective action  on $S^3$ from Abel-Plana formula.}
\label{ap1}

Let us see that the derivation of the effective action  on $S^3$ using the Abel-Plana formula (APF) gives the same results.
This is important because the absence of the topological  subleading  entropy in previous calculations \cite{emilio}
 introduced some confusion.
 The problem arises because of the presence of branch-point singularities, which make the application of the
 APF more subtle than usual.

 \medskip

The calculation is based on the application of the APF \cite{erd} to the function $f(z) = z^2\log(1-e^{-\beta z})$, which appeared in the computation of the effective action \eqref{eff3low} for the scalar field on the 3-sphere. We shall see that even when we are dealing with a function with branch-point singularities
on the imaginary axis, the APF can be applied provided they are integrable.

\begin{figure}[b]
  \centerline{\includegraphics[height=3cm]{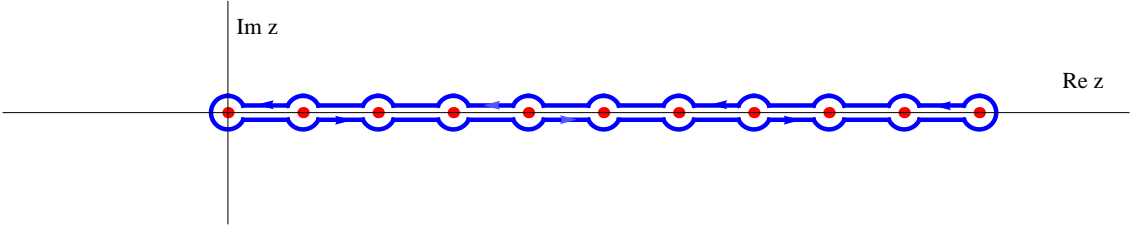}}
  \caption{\footnotesize{Initial contour of the Abel-Plana method.}} \label{abelplana}
\end{figure}

Let us  first analyze the derivation of the APF. From the argument principle we have that
\begin{equation}\nonumber
  \sum_{n=0}^{\infty} f(n) = \oint_{\mathcal{C}} dz \frac{f(z)}{e^{2\pi iz}-1}\, ,
\end{equation}
where $\mathcal{C}$ is a contour enclosing the values of $n$ in the sum --that is to say, surrounding the positive real axis, including the origin (see figure \ref{abelplana}).

Next, we exploit the fact that the functions are analytic in the complex right-half plane and behave properly at infinity to deform the contour $\mathcal{C}$ to the full imaginary axis (again, avoiding the origin) and rewrite the integrals there to obtain the APF in its usual form.

\smallskip

\begin{figure}[t]
  \centerline{\includegraphics[height=10 cm]{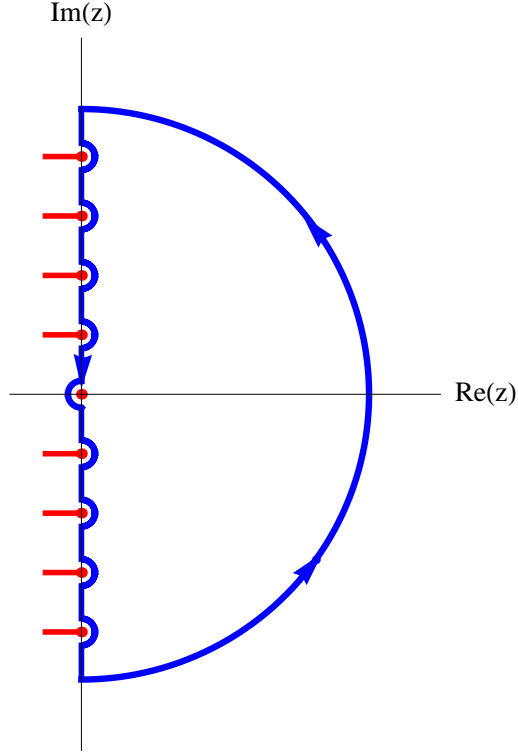}}
  \caption{\footnotesize{Deformed contour of the Abel-Plana method.}} \label{abelplana2}
\end{figure}

Now, as we already said, the function $f(z)$ we are interested in has branch-point singularities, located at the points $z_n=\frac{2\pi n}{\beta}i$ on the imaginary axis, where $n$ is an integer, so in deforming the original contour to the imaginary axis we have to avoid passing by this points, say, by using small semi-circumferences to the right of the axis (see figure \ref{abelplana2}). In principle, this might introduce some kind of modification to the APF. But we can easily see that the integral of $\frac{f(z)}{e^{2\pi iz}-1}$ over any of the semicircles centered at $z_n$ vanishes as its radius does, so the complete integral along this contour is equivalent to the integral over the whole imaginary axis, which leads to the usual APF
\begin{equation}\nonumber
  \sum_{n=0}^{\infty} f(n) = \int_0^{\infty} dx\, f(x)+  i\int_0^{\infty} dx\, \frac{f(ix)-f(-ix)}{e^{2\pi x}-1}\,\, ,
\end{equation}

The calculation of the first integral is straightforward. The result is
\beq\nonumber
 \int_0^{\infty} dk\, k^2 \log(1-e^{-\frac{\beta k}{a}})= -\frac{\pi^4 a^3}{45 \beta^3}
\eeq

In the second term  we have to deal with the logarithms with some care due to the changes of the arguments with $x$. The result is shown to be
\begin{eqnarray}\label{appri}
  S_{\mathrm{eff}, S^3} = -\frac{\pi^4 a^3}{45 \beta^3}+ \frac{\zeta(3)}{4\pi^2} + 2\pi \sum_{n=1}^{\infty} n \int_{{2n\pi a}/{\beta}}^{{2(n+1)\pi a}/{\beta}} dx\, \frac{x^2}{e^{2\pi x}-1}.
\end{eqnarray}

The summation term can be further simplified.
Using the geometric sum formula
\begin{equation}\nonumber
  \frac{1}{e^{{2\pi x}-1}} = e^{-2\pi x} \frac{1}{1-e^{-2\pi x}} = e^{-2\pi x} \sum_{k=0}^{\infty} e^{-2\pi kx}
\end{equation}
and interchanging  this new sum with the (finite) integral, we obtain the following result for this last term
\begin{align}
\nonumber
 -& 2\pi\sum_{l=1}^{\infty} l \sum_{k=1}^{\infty} \frac{e^{-2\pi kx}}{4\pi^3 k^3} (1 + 2\pi kx + 2\pi^2k^2x^2)\Bigr|^{{2(l+1)\pi}/{\beta}}_{{2l\pi}/{\beta}}
  \\ \nonumber =& -\frac{1}{4\pi^2} \sum_{l=1}^{\infty} l \sum_{k=1}^{\infty}\frac{1}{k^3}\left\{ e^{-\frac{4\pi^2(l+1)k}{\beta}} \left[2+2\frac{4\pi^2(l+1)k}{\beta}+\left(\frac{4\pi^2(l+1)k}{\beta}\right)^2\right]\right. -  \\ \nonumber &\hspace{3.5cm}- \left.e^{-\frac{4\pi^2lk}{\beta}} \left[2+2\frac{4\pi^2lk}{\beta}+\left(\frac{4\pi^2lk}{\beta}\right)^2\right]\right\}\,.
\end{align}

Now, we can change the index in the first term of the $l$-sum to obtain finally
\begin{equation}
 S_{\mathrm{eff}, S^3} =  -\frac{\pi^4 a^3}{45 \beta^3} + \frac{\zeta(3)}{4\pi^2} + \frac{1}{4\pi^2} \sum_{l,k=1}^{\infty} \frac{1}{k^3} e^{-\frac{4\pi^2lk}{\beta}} \left[ 2 + 2\frac{4\pi^2lk}{\beta} + \left( \frac{4\pi^2lk}{\beta}\right)^2 \right]\, ,
\end{equation}
 which coincides with the one obtained by zeta-function regularization (eq. \eqref{eff3}). The last two terms, that are missing in ref. \cite{emilio}, arise from the above-mentioned jumps in the argument of the complex logarithms.

\bigskip

{The Abel-Plana derivation of the topological subleading entropy shows that it can be related to the Stefan-Boltzmann leading term of the entropy of the
same conformal theory in one less dimension. Indeed, in this case entropy can also
be calculated from the Abel-Plana formula. The effective action
of a conformal scalar  on a  $S^2\times S^1$ space-time is  given by the expression \eqref{finite}
with $\lambda_k=\frac1{2a}(2k+1)$ and $d_k=2 k+1$, i.e.
\beq
\nonumber
S_{\rm eff}-\beta E_0=\sum_{k=1}^\infty (2 k +1) \log(1-e^{-\frac{\beta (2k +1)}{2 a}})\,=
\sum_{k=1}^\infty k  \log\frac{1-e^{-\frac{\beta k}{2 a} }}{(1-e^{- \frac{\beta k}{a}})^2}\, .
\eeq
Thus, the extensive term of this action is

\beq\nonumber
\int_0^\infty dk\,  k\, \log\frac{1-e^{-\frac{\beta k}{2 a} }}{(1-e^{- \frac{\beta k}{a}})^2}\,.
\eeq
The corresponding Stefan-Boltzmann contribution to the entropy  $S=(\beta\partial_\beta -1)S_{\rm eff}$ is then given by
\beq\nonumber
S_P= \beta \int_0^\infty dk\,  \frac{k^2}{2 a}\, \left[\frac1{e^{\frac{\beta k}{2 a}}-1}-\frac4{e^{\frac{\beta k}{a}}-1}\right]-\int_0^\infty dk\,  k\, \log\frac{1-e^{-\frac{\beta k}{2 a}}}{(1-e^{- \frac{\beta k}{ a}})^2}\,\,.
\eeq
or
\beq\nonumber
S_P= \frac{2 a^2}{\beta ^2}\int_0^\infty dk\,  \frac {k^2}{e^{ k}-1}-\frac{a^2}{\beta^2}\int_0^\infty dk\,  k\, \log\frac{1-e^{-\frac{ k}{2 }}}{(1-e^{-k})^2}\,\,.
\eeq
Now, scaling arguments show that
\beq\nonumber
S_P= \frac{3 a^2}{\beta^2}\int_0^\infty dk\, \frac {k^2}{e^{ k}-1}=  \frac{6\, \zeta(3) a^2}{\beta^2}\, ,
\eeq
and the corresponding entropy density is
\beq\label{enden}
s_{_P}=\frac{S_P}{4 \pi a^2}= \frac3{4 \pi \beta^2}\int_0^\infty dk\, \frac {k^2}{e^{ k}-1}= \frac{3 \,{\zeta(3)}}{2\pi \beta^2}\, .
\eeq
Since $s_{_P}$  does not depend on the radius of the sphere it does coincide with the entropy density
for the same theory on $\R^2\times S^1$ space-times. In fact, this Stefan-Boltzmann density contribution is universal and
the same for all compact manifolds.
On the other hand, the topological subleading entropy of the conformal massless scalar field theory on $S^3$ is given by (see \eqref{appri})
\beq
\label{topen}
S_{{\rm top}, S^3}= -\pi \int_0^\infty dk\, \frac {k^2}{e^{ 2 \pi k}-1}=- \frac1{4 \pi^2} {\zeta(3)}{}\, .
\eeq

The  interesting close relation between both quantities
\beq
\nonumber
S_{{\rm top}, S^3}= -\frac{\beta^2}{6\pi}  s_{P,S^2} \, ,
\eeq
 is due to their dependence on the same integral in formulas \eqref{enden} and \eqref{topen}, which
can be generalized for arbitrary dimensions thanks to the relation between both Abel-Plana formulas.}

\end{document}